\documentclass[runningheads]{llncs}
\usepackage{graphicx}
\usepackage{times}

\usepackage{soul}
\usepackage{url}
\usepackage[hidelinks]{hyperref}
\usepackage[utf8]{inputenc}
\usepackage[small]{caption}
\usepackage{graphicx}
\usepackage{amsmath}
\usepackage{booktabs}
\usepackage[utf8]{inputenc} % allow utf-8 input
\usepackage[T1]{fontenc}    % use 8-bit T1 fonts
\usepackage{booktabs}       % professional-quality tables
\usepackage{amsfonts}       % blackboard math symbols
\usepackage{nicefrac}       % compact symbols for 1/2, etc.
\usepackage{microtype}      % microtypography
\usepackage{lipsum}
\usepackage{multirow}

\begin{document}
\title{A Large-scale Industrial and Professional Occupation Dataset}
\titlerunning{Industrial and Professional Occupation Dataset}
\author{
Junhua Liu\inst{1} \and 
Yung Chuen Ng\inst{2} \and 
Kwan Hui Lim\inst{1}
}

\authorrunning{J. Liu et al.}

\institute{
 Singapore University of Technology and Design\\
 \email{junhua\_liu@mymail.sutd.edu.sg, kwanhui\_lim@sutd.edu.sg} \and
 National University of Singapore\\
 \email{e0201912@u.nus.edu}}

\maketitle              % typeset the header of the contribution
\begin{abstract}
There has been growing interest in utilizing occupational data mining and analysis. In today's job market, occupational data mining and analysis is growing in importance as it enables companies to predict employee turnover, model career trajectories, screen through resumes and perform other human resource tasks. A key requirement to facilitate these tasks is the need for an occupation-related dataset. However, most research use proprietary datasets or do not make their dataset publicly available, thus impeding development in this area. To solve this issue, we present the Industrial and Professional Occupation Dataset (IPOD), which comprises 192k job titles belonging to 56k LinkedIn users. In addition to making IPOD publicly available, we also: (i) manually annotate each job title with its associated level of seniority, domain of work and location; and (ii) provide embedding for job titles and discuss various use cases. This dataset is publicly available at https://github.com/junhua/ipod.

\end{abstract}

%=========================%
\section{Introduction}
\label{sec:introduction}

Occupational data mining and analysis is a popular research topic in recent years. There are many lines of research within occupational data mining and analysis, including predicting employee turnover~\cite{yang2018one,zhao2018employee}, modelling and predicting career trajectories~\cite{liu2016fortune,mimno2008modeling}, predicting employee behaviors ~\cite{chen2012mining,cetintas2011identifying} and various others. A common requirement among these works is the need for a occupation-related dataset, which could be derived from professional networking sites (e.g., LinkedIn), scraped from online resumes or other sources. However, most of these datasets are not publicly available, thus impeding future research in this area. To address this problem, we curate and make publicly available the Industrial and Professional Occupation Dataset (IPOD), which comprises 192,295 job titles/positions belonging to 56,648 users on LinkedIn. To the best of our knowledge, IPOD is the largest publicly available occupation-related dataset. This dataset will be useful for researchers and industry practitioners who are interested in occupational data mining and analysis.

\subsection{Related Works}
There has been numerous works in recent years that utilize related datasets. We performed a comprehensive literature review of papers since 2008 and identified 15 related works utilizing such datasets.\footnote{Due to space constraints, we are unable to list all 15 papers but details are provided in~\cite{liu2019ipod}.} However, out of these 15 works, only one~\cite{james2018prediction} made their dataset publicly available. This dataset comprises the names of affiliations of physics scientists without their job titles, whereas our dataset comprises the job titles across the broader industry.

%=========================%
\section{Dataset Description}
\label{sectDataset}

\begin{figure*}[t]
\centering
  \includegraphics[width=\textwidth, trim=5cm 0 5cm 0]{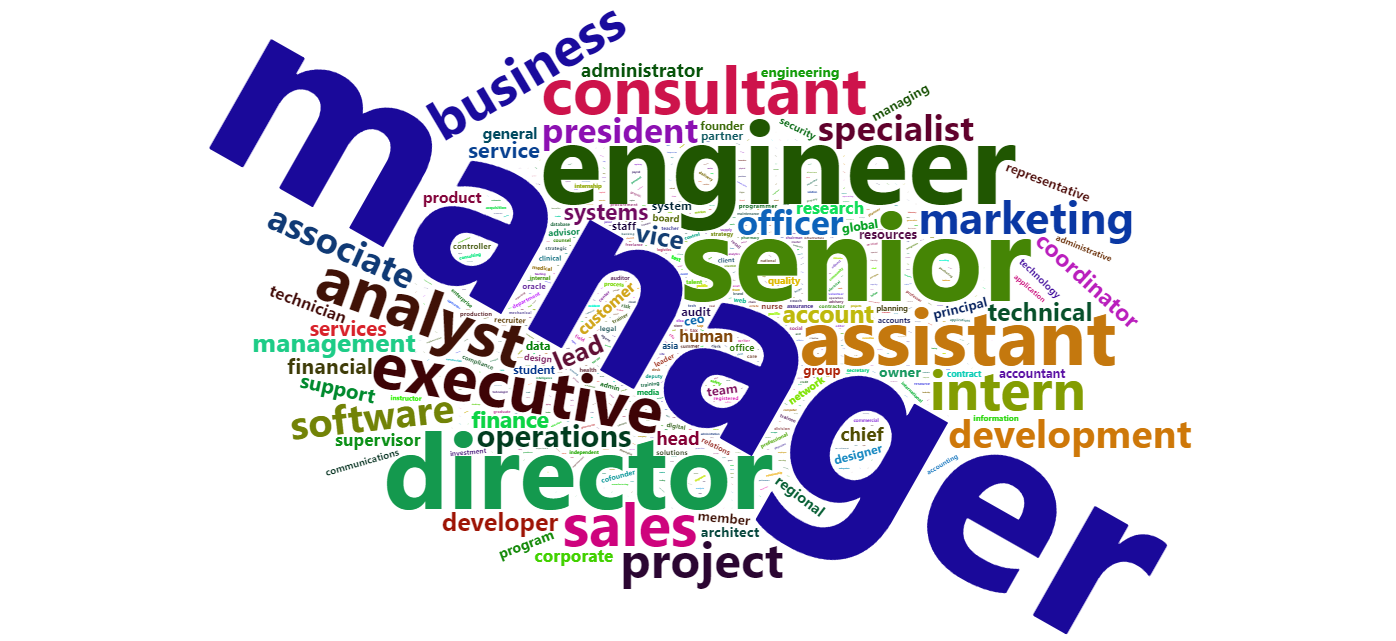}
  \caption{Wordcloud of the job titles in IPOD.}
  \label{fig:jobcloud}
\end{figure*}

The IPOD dataset comprises a total of 192,295 job titles/positions that were crawled from the LinkedIn profiles of 56,648 users. $56.7\%$ and $43.3\%$ of these profiles were from the United States and Asia, respectively. Figure~\ref{fig:jobcloud} shows a wordcloud of the job titles in our dataset.

\begin{table}[h]
\centering
\begin{tabular}{ll}
\hline
Responsibility (RES) & \begin{tabular}[c]{@{}l@{}}Managerial level: \textit{manager, director, president, etc}\\ Operational role: \textit{technician, engineer, accountant, etc}\\ Seniority: \textit{junior, senior, chief, etc}\end{tabular} \\ \hline

Function (FUN) & \begin{tabular}[c]{@{}l@{}}Departments: \textit{marketing, operations, finance, etc}\\ Scope: \textit{enterprise, national, international, etc}\\ Content: \textit{security, education, r\&d, etc}\end{tabular} \\ \hline

Location (LOC) & \begin{tabular}[c]{@{}l@{}}Regions: \textit{Asia, Europe, SEA, etc}\\ Countries/States/Cities: \textit{Singapore, California, Melbourne, etc}\end{tabular} \\ \hline
\end{tabular}%
\caption{Examples of tags associated with job titles.}
\label{tab:NEsample}
\end{table}

{\bf Structure of Dataset}. The job titles represent the various positions held by a person, particularly the level of seniority, domain of work and location. As shown in Table~\ref{tab:NEsample}, we label each job title with the following tags:
\begin{itemize}
\item Responsibility (RES). This tag indicates the level of responsibility associated with a job. This RES tag can be further divided into managerial level, operational role and seniority. For example, ``Senior Director'' corresponds to seniority and managerial level, respectively, while ``Junior Technician'' corresponds to seniority and operational role, respectively.
\item Function (FUN). This tag indicates the typical business functions in organizations. Similarly, the FUN tag can be further divided into the department, scope of work and content.
\item Location (LOC). This tag indicates the geographic locality that the job title is responsible for, which could be at the scope of a region (e.g., Europe) or a smaller area like a city, state or country (e.g. Singapore).
\item Others (O). This tag is for any other tokens which do not fall into the earlier three categories.
\end{itemize}

{\bf Annotation Process}. Our labelling is performed by three annotators who are highly experienced with such job titles and the tagging task, namely a Human Resource personnel, senior recruiter and business owner. From our corpus of job titles, we extracted 1,500 tokens of the most frequently occurring uni-grams which are labelled by the three annotators. The Inter-Rater Reliability scores based on two inter-annotator agreements show a score of 0.853 for Percentage Agreement~\cite{viera2005understanding} and 0.778 for Cohen’s Kappa~\cite{artstein2008inter}, which represents a \textit{Strong} level of agreement among annotators. We also observe that 77.9\% of labelled tags are agreed by all three annotators, 22.1\% are between two annotators, while there are no cases where all three annotators disagree on a label.

%=========================%
\section{Use Cases}

In this section, we briefly describe various possible use cases of our IPOD dataset. For a detailed write-up of the algorithms in these use cases, we refer readers to~\cite{liu2019ipod} that is currently under review at the Applied Data Science track of ECML-PKDD 2020.

\subsection{Embedding for Job Titles}
One use case of our IPOD dataset is to generate embedding for the job titles, which will enable us to perform various occupational data mining tasks. For this purpose, we also develop and release an embedding for job titles, \textit{Title2vec}, which we generate using a deep bidirectional language model (biLM) that is fine-tuned from pre-trained ELMo embeddings on a large text corpus~\cite{peters2018deep}. \textit{Title2vec} is useful for numerous tasks, such as understanding similar job titles across different companies, or as the input to career trajectory prediction problems, job turnover prediction problem and other similar tasks.

\subsection{Occupational Named Entity Recognition}
Another use case of IPOD is for the Occupational Named Entity Recognition (NER) task. Traditional NER tasks uses general tags such as \textbf{PER}son, \textbf{ORG}anization, etc, whereas in our occupational NER task, we have more specialized and domain-specific tags such as \textbf{RES}ponsibility, \textbf{FUN}ction and \textbf{LOC}ation as previously described in Section~\ref{sectDataset}. With the ever-changing industry landscape and cultural difference between international workplaces, this occupational NER tasks allow us to better understand the profile of emerging job titles and identify similar job positions across different countries.

%=========================%
\section{Conclusion}

In this paper, we present the IPOD dataset for occupational data mining and analysis tasks, comprising the job titles, manually annotated tags and a \textit{Title2vec} embedding for the job titles. This dataset comprises 192k job titles belonging to 56k LinkedIn users. To the best of our knowledge, IPOD is the largest publicly available dataset that contains occupational information about the general industry.

%=========================%
\section{Acknowledgement}
This research is funded in part by the Singapore University of Technology and Design under grant SRG-ISTD-2018-140.

% https://www.github.com/junhua/ipod
\bibliographystyle{splncs04}
\bibliography{ref}
\end{document}